\documentclass[prd,twocolumn,showpacs,preprintnumbers,amsmath,amssymb,epsfig,widetext,floatfix,superscriptaddress]{revtex4}

\usepackage{graphicx}
\usepackage{dcolumn}
\usepackage{bm}
\usepackage{color}
\usepackage{natbib}
\usepackage{rotating}
\usepackage{times}

\begin{document}

\title{Intensity Scintillation and Astronomical Quantum Observation}

\author{Jiang Dong}
\email{djcosmic@gmail.com}
\affiliation{YunNan Astronomical Observatory,NAOs,CAS;\\
396,JinMa Town,GuanDu District,KunMing, YunNan 650011,China}
\affiliation{Scientific and application center of lunar and deepspace exploration, 
NAOs, China;\\
djcosmic@gmail.com}
\date{\today}

\begin{abstract}
\baselineskip 11pt
Holography is 3D imaging which can record intensity and phase at the same time.
The importance of construct hologram is 
holographic recording and wavefront reconstruction.  
It is surprised that holography be discovered 
in study interstellar scintillation for pulsar provide a coherent
light source recently.
I think that is speckle hologram and 
speckle interference(i.e. intensity interference),
and use modern technique which include phased array,CCD, 
digital signal processing and supercomputer can achieve that 
digital and computer holography from radio to X-ray astronomy.
This means we can use it to image the universe
and beyond the limited of telescope for cosmos
provide much coherent light from pulsar,maser, black hole to 21cm recombination line.
It gives a probe to the medium of near the black hole et al.
From those coherent light sources in the sky,
we can uncover one different universe that 
through astronomical quantum observation which use intensity interference. 
\end{abstract}
\maketitle

\section{Introduction} \label{sec:intro}

After the invention of the telescope in 1608 in the Netherlands, 
by Hans Lippershey, Zacharias Janssen and Jacob Metius, 
it was the refinement of the telescope the following year by Galileo
that triggered one of the greatest scientific revolutions of all time. 
Now astronomical observation has extend to all waveband of electromagnetic wave
from radio to $\gamma$-ray, cosmic rays, neutrino and gravitational wave.
In usually,astronomer just measures celestial signal 
in the spatial or temporal that involves
imaging, spectroscopy, polarimetry and variation in time.

The hologram is invented by Gabor,D in 1947,
which base Bragg's wavefronts reconstruction in microscopy and Zernike's phase contrast.
It originates in classic wave optic and fourier optic
but thrives for maser and laser development that
provide a excellent coherent referenced light.
Now holography has been widely applied
in many areas which include 3D displays, optical memories,
diffuser display screen,
nodestructive measurement and testing et al.\cite{lgr02}.

It is surprised that holography be discovered 
pulsar astronomy recently\cite{wkss08}.
I think it is speckle hologram for pulsar provide a coherent light source.
Speckle is a random intensity pattern produced by 
the mutual interference of a set of wavefronts. 
This phenomenon has been investigated by scientists since the time of Newton
to explain that twinkle stars, 
but speckles have come into prominence 
since the invention of the laser and have now found a variety of applications
which include optic coherence tomography,optical projection display,
projection microlithography, laser radar(LADAR),
Imaging Coherent Optical Radar (ICHOR) and 
Infrared Coherent infrared imaging camera (CIRIC) et al\cite{gjw06}.
The speckle effect as a result of the interference of many waves,
having different phases, which add together to give a resultant wave
whose amplitude and intensity varies randomly. 
It was considered to be a severe drawback in using lasers to illuminate objects,
particularly in holographic imaging because of the grainy image produced.
It was later realized that speckle patterns could carry information 
about the object's surface deformations, 
and this effect is exploited in holographic interferometry
that a technique enables static and dynamic displacements of objects 
with optically rough surfaces to be measured to optical interferometric precision. 
These measurements can be applied to stress, strain and vibration analysis, 
as well as to non-destructive testing. 
It can also be used to detect optical path length variations in transparent media, 
which enables, for example, fluid flow to be visualised and analysised.
And in astronomy, it widely is studied in stellar speckle of
optical\cite{gjw06,lgr02}.

\section{Intensity Scintillation and Speckle Holography
in the Universe} \label{sec:IS}

Astronomical interferometer has two types in general. 
One is spatial coherence like Young’s ‘two-slit experiment’, 
the celestial signal interfere in the same phase,
for example VLBA,VLA,EVN et al.
Another is temporal coherence(i.e. Intensity Interferometer)
like Michelson interferometer,
the celestial signal interfere in the intensity,
for instance Narrabri Stellar Intensity Interferometer\cite{hbt56}.
Speckle interferometry is intensity interference.
In 1970, Labeyrie, A. showed that information could be obtained 
about the high-resolution structure of the object
from the speckle patterns using fourier analysis (speckle
interferometry)\cite{l70}. 
Then it was first demonstrated in astronomical imaging in
1972\cite{gls72}.
If a reference point source is available within the iso-planatic patch, 
it is used as a key to reconstruct the target in the same way 
as a reference coherent beam is employed in holographic reconstruction 
\cite{s02}.
In the 1980s the methods were developed 
which allowed images to be reconstructed interferometrically 
from these speckle patterns but speckle hologram be advised in 1973\cite{bgn73}.
It is widely employed both in the visible, as well
as in the infrared (IR) bands at telescopes 
to decipher diffraction-limited information.
Since its advantage, it is well in double star\cite{bm+08} 
and solar granules\cite{hb73}.  
Then differential speckle interferometry, speckles and shadow bands,
speckle spectroscopy,speckle polarimetry and speckle imaging of extended objects
be deeply studied in optic astronomy\cite{s02}.

Researching interstellar scintillation, Stinebring et al. 
produced two-dimensional power spectra from the dynamic spectra.
The crisscross pattern in the dynamic spectrum
shows up in the secondary spectrum as two parabolic features,
curving away from the conjugate time axis\cite{smc+01}.
Fig.1 is the dynamic and secondary spectra of PSR B1133+16, 
shows multiple scintillation arcs on occasion\cite{crsc06}.
The secondary spectra produced by the auto-correlation and 
fourier transformation, Rickett give the relationships between primary,
secondary spectra, scattered brightness distribution, pulse shape, and
the second moments of the field\cite{r06}.
So the secondary spectra is the result of the intensity interfere 
in the data of different waveband (channels), 
and is vary in the fourth order of electromagnetic field.
The explain of parabolic arc use a stationary screen\cite{crsc06} 
or phase \cite{wmsz04},and result of anisotropic scattering. 
The integration time is important in determining 
whether the scattered brightness(image-intensity) produced 
by fourier transforming the windowed amplitudes, 
and taking the squared modulus of the result, 
as discussed by Narayan \& Goodman \cite{ng89}. 
An instantaneous angular spectrum is a single realization of a random process. 
Thus, it will exhibit "speckle" (i.e. the components of the angular spectrum 
are statistically independent) having an exponential distribution. 
This is the "snapshot" scattered image, 
as described by Goodman \& Narayan \cite{gn89} for a very short integration 
from a filled aperture telescope. 
The speckle is reduced when the integration time is longer than 
the diffractive scintillation time.
The similar data reduce process give in optical astronomy 
by Dravins et al\cite{dlmy97a} for study 
atmospheric intensity scintillation of stars.
They have the same result in the dependence 
on wavelength(frequency)\cite{dlmy97b,hsb+03}.
The similar analysis is used to compact structures in the galaxy\cite{hsa+05}
and structure of "flying shadow"\cite{dlmy98}.
The research of optic \cite{dlmy98} shows that integration time,
large apertures and coherent light source is important in produce arclets.

Comparing between optic and radio,I think interstellar holography 
in pulsar astronomy\cite{wkss08} is speckle holography like in optic\cite{l70,bgn73,s02}.
Cordes has given intensity cross-correlation methods for inferring source sizes.
Intensity interferometry in the radio context is compared to the optical intensity
interferometry of Hanbury-Brown and Twiss\cite{c00}.

Like pursuing "untwinkling star" in optic which gives 
active systems in Second-Order adaptive optics that correcting
not only the phase due to the angular tilt of the wavefront, 
but also amplitude effects from its curvature, 
pulsar astronomer also correct for interstellar scattering delay 
in pulsar timing for detect gravitational waves use pulsar\cite{hs08}. 
Another interesting thing is to use speckle hologram 
resolve pulsar magnetosphere,
give structure of emission area, define the radiation mechanism use 
the character of monochromatic in hologram. 

Double pulsar system PSR J0737-3039 is a special case, 
which can provide two coherent light,
it is a pity that they still not find arc in the observation\cite{cmr+05}.
Another unusual case is pulsar B1641-45 and OH maser
system\cite{wjks06},
it also has multi-coherent light sources. 
If speckle holography is discovered in the similar system, 
it will give strong limit to interstellar medium(ISM).
Whether or not exist modulate between multi-coherent light system like
in lab\cite{b82} is interesting things too.

The phase space for all kinds of radio transients show in Fig.2. 
Maser is strong coherent source\cite{e92} in radio astronomy 
from the Milk Way to the extragalaxy\cite{k05}.
Using the same method, we must be gain the speckle effect in it.
It will provide the more information of star formation area or ISM.
The electron-cyclotron maser as coherent radiation be apply to explain 
the radio emissions from magnetized planets, auroral kilometric radiation,
 Jupiter radio bursts,solar radio bursts/spikes,coherent radiation
from stars and coherent radiation from Blazar jets\cite{t06}.
That means speckle interference 
(intensity interferometer) have extensive application.

Coles and Rickett extend this way to solar radio observation,
it provides a precise measurement of the flow speed 
and a two dimensional image of the scattered radio waves,
from a broadband measurement with a single antenna\cite{cr06}.
The process like use speckle to detect optical path length variations 
in transparent media, for example, fluid flow to be visualised and analysised.
The interesting thing is study solar use multiwavelenth speckle
that involve radio, optical and  X-ray\cite{kbc+08}.

Intensity scintillation or speckle holography in X-ray
is significant for it means we can have more information which near
the black hole\cite{rm07}, center of AGN \cite{hk06}and 
stellar interior involve solar\cite{kbc+08}.
Fig.3 is a artistic graphic expression of AGN model,
X-ray come from jet in synchrotron et al.\cite{hk06},
it also provide x-ray free-electron lasers.
Recombination radiation in usually as X-Ray laser emission,
it also can be produced in AGN.
Until now, we confirm black hole all use indirect way like
mass-luminosity relation.
If we can record intensity scintillation in X-ray,
we will have chance image black hole almost directly 
for it will offer exclusive news from near the horizon.

\section{Digital Recording and Numerical Reconstruction 
of Wave Fields in Holography} \label{sec:QO}

In recent twenty years, the hologram has huge development,
especially in optical 3D display et al.\cite{lgr02}. 
Digital holography is use CCD, computer et al achieve interference record.
Computer generated hologram simulate the process of wavefront reconstruct
use computing.

With CCD apply from infrared, optic to soft X-ray,it can easily achieve 
interference record.
In radio astronomy, digital radio include phased array antenna(feed), 
high speed A/D, digital signal processing and supercomputer et al.
also benefit in hologram record.
Software holography has been use to removing 
the assumption of a single complex gain per antenna 
and allowing more complete descriptions of the instrumental 
and atmospheric conditions that the similarity with 
holographic mapping of radio antenna surfaces\cite{mm08}.
It is first order adapt optic in radio astronomy for 
it only correct in phase space.

Now X Ray Holographic can realize in hard X-ray use free-electron
lasers\cite{sca+08},
they are able to resolve atomic distances, 
and can give the 3-D arrangement of atoms around a selected element. 
Therefore, hard X-ray holography has potential applications at structural
research in chemistry, biomolecular imaging of biology, and physics\cite{ft03}.
In astronomy, X-ray interferometry be advised recently,
but most of telescope base the complex optical design\cite{c03},
one simple path is intensity interference use 
in software like speckle interference in optic and radio.
We just need improve the sample rate of data in time resolution,
then use fourier transform to interference in software.

The simulation of speckle procession that is numerical reconstruction 
of wave fields is important for it give clue to 
understand the result of interference, like walker et al. 
make Monte Carlo simulations of snapshot secondary spectra 
in explain arclet\cite{wmsz04} when imaging black hole or solar.

\section{Beyond Galileo - Astronomical Quantum Observation} \label{sec:AQO}

From Ptolemy times, we aways to give a map of universe.
After Galileo use optical telescope point to the sky, 
we begin to realize macro world. 
In 1908, Lippmann invent colour photographically 
based on the phenomenon of interference(noble in 1908),
so we can record in mesoscopic. 
Microscope has gigantic development with physical scientist 
understand quantum phenomenon.

Almost all of four hundreds years, 
astronomer measure either the spatial or temporal [first-order] 
coherence of light which from radio to $\gamma$ ray. 
The spatial coherence(in various directions, and for different angular
extents on the sky) are studied in imaging devices that include 
cameras, interferometers et al. 
All spectrally analyzing devices measure aspects of the temporal coherence
that involve different temporal/spectral resolution, 
and in the different polarizations.

The best-known non-classical property of light, 
first measured by Hanbury Brown and Twiss in those experiments that 
led to the astronomical intensity interferometer\cite{hbt56}.
For explain the result of Hanbury Brown in optical intensity interferometer
and the properties of laser beams in labs
that use quantum mechanical theory,
the quantum optic (coherent optic) be developed 
by theory physicist that include Glauber et al.\cite{g06}.
In labs, the statistics optic be use to measure 
the character of laser\cite{gjw06}.

Now photonics and quantum optics, studying individual photons, and photon streams. 
One photon stream has further degrees of freedom, 
such as the temporal statistics of photon arrival times, 
giving a measure of ordering (entropy) within the photon-stream, 
and its possible deviations from "randomness". 
Such properties are reflected in the second-(and higher-) order coherence of light, 
observable as correlations between pairs (or a greater number) of photons. 
Those can be complex, carrying information beyond imaging,spectroscopy, or polarimetry. 
Different physical processes in the generation of light 
may cause quantum-statistical differences 
(e.g., different degrees of photon bunching in time) between 
light with otherwise identical spectrum, polarization, intensity, etc., 
and studies of such nonclassical properties of light 
are actively pursued in laboratory optics\cite{d08,g06}. 
Since almost all astronomy is based upon the interpretation of subtleties in
the light from astronomical sources, 
quantum optics appears to have the potential of becoming another information channel 
from the Universe, fundamentally different from imaging and spectroscopy\cite{d08}.

In optical astronomy, just a few star that include Eta Carinae and MWC 349
et al. need laser theory to explain for most of radiation process
origin from local thermodynamic equilibrium (LTE).
If we study intensity scintillation in optic, 
that will need highest time resolution instrument\cite{d08}.
But in radio and X-ray astronomy, full of non-LTE process,
cyclotron radiation, synchrotron radiation and recombination radiation
et al come from pulsar, maser, AGN(black hole) and X-ray binary et al. 
which provide best coherent light sources.
Cosmic turbulence which from interstellar medium to the
circumambience of black hole will imprint with wave(or photon stream)
and produce the propagation effects.
That all must be with quantum optics.
Some unknown result of Stinebring et al.\cite{smc+01,hsb+03,hsa+05,crsc06} 
also need quantum mechanical theory to explain
for they do intensity interference(interference in different channel data)in data process 
and pulsar radiation may be take place the speckle effect 
that lead by interstellar medium like laser in labs.

Now cartography must be improved for abundant coherent light sources,
quantum astronomy will give different information from traditional
observation(single dish, interferometry and synthesis) in phase space
like quantum optic in labs.
Astronomical quantum observation also can provide different phenomena
with labs for many source is so strong that can't be simulated in labs
and some surrounding only have in the cosmos.

The first light of the universe be thought will observed use large
antenna array like SKA, LOFAR, LWA, MWA and PAPER et al.
I think use the intensity interference may be easily have the result 
to give a map of 21 cm emission for the dissimilar character in
coherent of the recombination of hydrogen.
In traditional, interferometry and synthesis in radio astronomy
is spatial coherence that the photon(wave) have the same phase(wave front).
Intensity interferometer use to measure angular width by Hanbury Brown
et al. in early radio astronomy.
Radio astronomer abandoned it for its relative lack of
sensitivity\cite{tms01}. 
The 21cm signal is smaller than the observed photon energy,
the weak signal is accompanied by a flood of foreground photons 
with a high occupation number (involving the synchrotron Galactic emission 
and the cosmic microwave background). 
Although astronomer want to use high order polynomial filter 
to separate foregrounds, instrumental response and other
non-astrophysical effects (e.g. ionosphere, RFIs, etc.). 
The signal photons are not individually distinguishable in spatial coherence, 
the combined signal+foreground population of photons has 
a high occupation number in the phase space\cite{l08}.
The 21cm signal may be produce recombination lines masers\cite{sn97},
even if not\cite{dl08},
it will easiest gain the result use intensity interferometer than 
phase interferometer for most of signal that include foregrounds, 
instrumental response and other non-astrophysical effects, 
is non-coherent or have different character in coherence 
for example synchrotron Galactic emission.


In X-ray astronomy, XMM, Chandra et al. single dish telescope 
have launched, if we can give a intensity interference in software
like Stinebring et al., it must be have new result in quantum astronomy.
Not only simulation in computer is important but also
simulation in lab is integrant if we use the speckle effect to image
black hole. Now laboratory X-Ray astrophysics almost study spectrum
like laboratory microwave spectrum to molecular astronomy\cite{b03}. 
The turbulence in plasma can be simulated in labs now\cite{jbsg06}.
If we can use X-ray laser to produce a radiography to it like 
use dragon-I, DARHT-I and  AIRIX diagnose bomb,
that will benefit to we make sure that 
the photograph whether come from black hole directly.

\vspace{1.cm}

\noindent {\it Acknowledgments}: DJ thanks the inventor of internet.
\hfill\vfill

\bibliography{holobib}
 
\clearpage

\begin{figure}[tp]
\begin{center}
\includegraphics[width=3.4in]{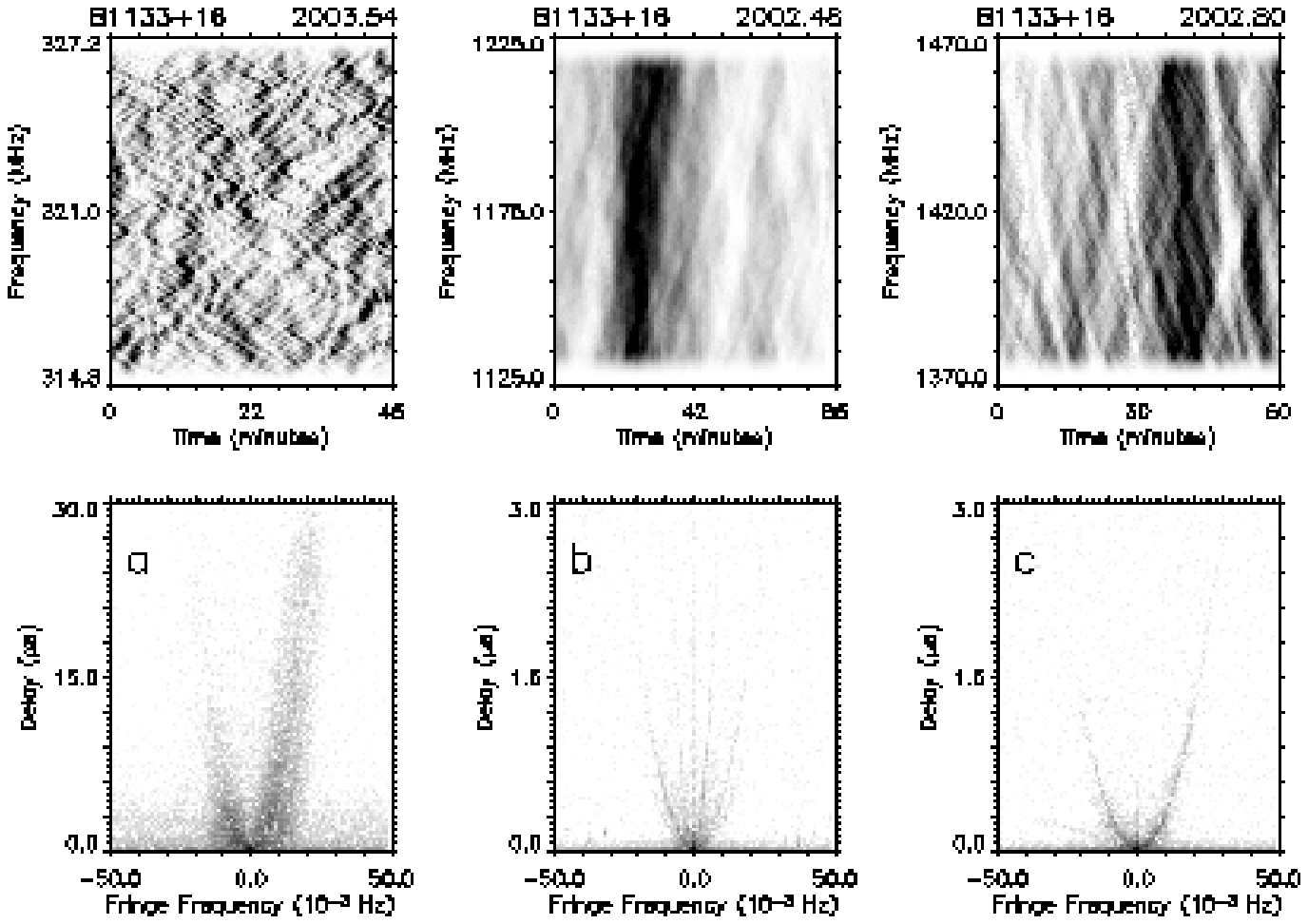}
\caption{``Secondary spectrum'' (or
``delay-Doppler''diagram) for an axisymmetric image with a Gaussian
intensity profile. The axes are expressed in units of $\theta_o=1/ks_o$,
for $q$, and $\theta_o^2$ for $p$. 
Contour levels are set at intervals of $3$~dB, and the lowest (outermost)
contour is at $-60$~dB relative to $P(0.1,0.1)$.
PSR B1133+16, shows multiple scintillation arcs on occasion. 
The broad, asymmetric power distribution in (a) has numerous arclets
at 321 MHz. Panels (b) and (c) are at frequencies above 1 GHz. 
Panel (b) shows two clear arcs (along with a horizontal line at ft due
to narrowband radio frequency interference 
and the sidelobe response of power near the origin). 
Four months later (panel [c]), only the outer of these two arcs, 
widened by the $\alpha \propto \nu^{-2}$ scaling, is visible. 
The gray scale for the dynamic spectrum is linear in flux density. 
For the secondary spectrum, the logarithm of the power is plotted, 
and the gray scale extends from the noise floor (white) to 5.5 decades above it
(black). The data were obtained from the Arecibo Observatory\cite{crsc06}.}
\label{fig:arclet}
\end{center}
\end{figure}

\begin{figure}[tp]
\begin{center}
\includegraphics[width=3.4in]{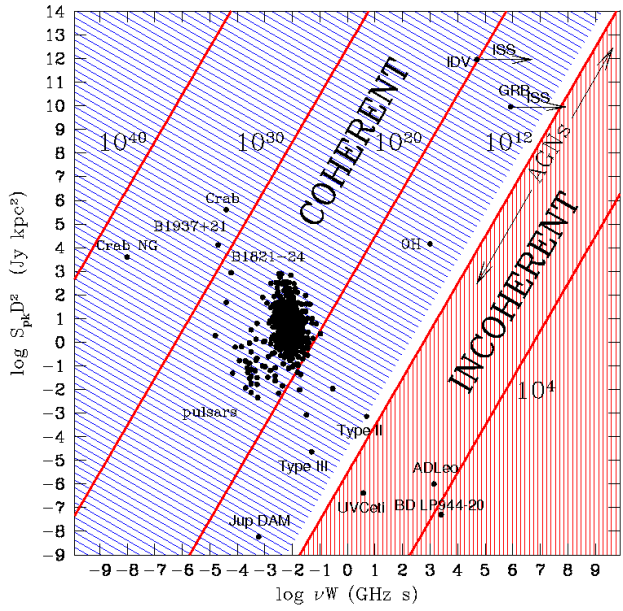}
\caption{The phase space for radio transients.  The abscissa is the
product of the emission frequency $\nu$ and transient duration or pulse
width~$W$.  The ordinate is the product of the observed flux
density~$S$ and square of the distance~$D^2$. The sloping
lines are labelled by constant brightness temperature\cite{clm04}.}
\label{fig:ps}
\end{center}
\end{figure}

\end{document}